\documentclass[aps,prd,showpacs,showkeys,preprint]{revtex4}
\usepackage{amsmath,amsthm,amsfonts,amssymb}
\usepackage[mathcal]{eucal}
\usepackage{mathrsfs}
\usepackage{graphicx}
\usepackage{dcolumn}
\usepackage{latexsym}
\usepackage{epsfig,color}
\usepackage{bm}
\usepackage[all]{xy}

\usepackage[utf8]{inputenc} 

\newcommand{\ie}{\begin{equation}}
\newcommand{\fe}{\end{equation}}

\def\text#1{\mbox{#1}}

\begin{document}

\title{{Gauge field localization on the Bloch Brane}}
\author{W. T. Cruz$^{a}$, Aristeu R. P. Lima$^b$ and C. A. S. Almeida$^{c}$}

\affiliation{$^a$Instituto Federal de Educa\c{c}\~{a}o, Ci\^{e}ncia e Tecnologia do Cear\'{a} (IFCE), Campus Juazeiro do Norte, 63040-000 Juazeiro do Norte-Cear\'{a}-Brazil}
\affiliation{$^b$Universidade da Integra\c{c}\~{a}o Internacional da Lusofonia Afro-Brasileira, Campus da Liberdade, 62790-000 Reden\c{c}\~{a}o-Cear\'{a}-Brazil}
\affiliation{$^c$Departamento de F\'{i}sica - Universidade Federal do Cear\'{a} \\ C.P. 6030, 60455-760 Fortaleza-Cear\'{a}-Brazil}

\begin{abstract}
We present new results on the localization of gauge fields in thick brane models. The four-dimensional observable universe is considered to be a topological defect which is generated by two scalar fields coupled with gravity embedded in a five-dimensional space-time. Like other thick brane models, the Bloch brane setup is not capable of supporting gauge field localization. In order to circumvent this problem we include the dilaton field and, as a result, obtain normalizable solutions. In addition, by writing the equations of motion as Schroedinger-like equations, we have found new kinds of resonances in the massive spectrum which appear also for heavy modes. This is in sharp contrast to what is commonly found in previous analog studies, where only light modes become resonant. At specific energies, the wave solutions exhibit very high amplitudes within the membrane. The influence of the dilaton coupling and of the internal structure on the resonant modes are also discussed.
\end{abstract}

\pacs{11.10.Kk, 11.27.+d, 04.50.-h, 12.60.-i}

\keywords{Field theories in higher dimensions, Braneworlds, Gauge vector field localization.}
\maketitle

\section{Introduction}
In brane-world models the observable universe is represented by a four-dimensional membrane embedded in a higher-dimensional space-time. In this context finding scenarios of membranes which mimic consistently the features of the Standard Model (SM) turns out to be an important problem. Such ideas have appeared as alternatives to solve the so called hierarchy problem and the the cosmological constant problem of the SM \cite{RS,add}. An important attempt to find possible solutions to the hierarchy problem is represented by the Randall-Sundrum (RS) model. However, that model does not favor the localization of vector gauge fields, an important ingredient in order to construct the SM in the membrane \cite{davo}.

Brane-world models have the property of allowing for the localization of several different types of fields, both bosonic and fermionic ones. Therefore, the study of topological defects in the context of warped space-times has been given much attention recently. For example, domain walls as extended defects in field theory have been used in high energy physics for representing brane scenarios with extra dimensions \cite{rub,vis}. The Bloch walls are interfaces that host internal structure which could be seen as chiral interfaces \cite{wall}. In the context of extra dimensions, the Bloch brane is constructed in a $(4,1)$ model of two scalar fields coupled with gravity \cite{dionisio}. Such brane model is generated dynamically, has internal structure and the asymptotic bulk metric is a slice of an five-dimensional anti-de Sitter (AdS) space, denoted by $AdS_5$.

The localization of gravity \cite{dionisio} and of fermions \cite{castro, carlos} has been considered in the context of Bloch brane recently, but a corresponding study for gauge field is still missing in the literature, to the best of our knowledge. Knowing the importance of having gauge field localized to guarantee the consistence of the model, we are motivated to study its behavior on the Bloch brane. We pay special attention to the search for resonant modes in the massive spectrum because such structures will give us details about the coupling of these modes with matter on the brane.

Some previous articles have considered the study of localization of gauge fields on membranes \cite{thickbrane1, thickbrane2, thickbrane3, thickbrane4, thickbrane5}. Specifically in the work of Guerrero et. al. \cite{thickbrane1} the brane thickness and spacetime curvature effects  on the localization of gauge fields zero modes on a brane via kinetic terms induced by localized fermions were considered. An effective action approach has also been considered in the context of gauge field localization \cite{thickbrane2}. Moreover, Chumbes et. al. \cite{thickbrane3} investigated the features of the gauge field zero mode on a thick brane generated by the coupling of a scalar field and gravity. Single brane localization, where a gravitational mechanism was used to localize gauge bosons and gravitons \cite{thickbrane4}, as well as multi-brane localization \cite{thickbrane5} have also been investigated.

In this paper we focus on the behavior of the gauge field on a thick brane with internal structure generated by two scalar fields. In order to achieve a zero mode localized we introduce a third scalar field, the dilaton, that will couple directly to the gauge field. We have analysed the massive spectrum searching for resonant modes with large life-times. The influence of the dilaton coupling and the brane thickness on the life-times of the resonant structures is also discussed.

This work is organized as follows. In Sec. \ref{rev_bb_sce}, we briefly review the Bloch brane scenario. In Sec. \ref{gau_f_ze_mode} we study the gauge field zero modes provided by the previous setup. Sec. \ref{dil_bb} is devoted to showing how the introduction of the dilaton field leads to the localization of the gauge field. The following section, Sec. \ref{mass_mode_res} is dedicated to studying the behavior of Kaluza-Klein (KK) modes and resonances. Finally, we present our conclusions and perspectives in Sec. \ref{conclu}.

\section{Reviewing the Bloch Brane scenario \label{rev_bb_sce}}

Initially, in order to set the stage for studying the localization of gauge fields in the present context, let us briefly review the thick brane scenario. The basic Bloch brane background was previously studied in references \cite{dionisio,adauto,carlos,castro}. This model consists of two real scalar fields coupled to gravity and has as a starting point the following action
\begin{equation}
S=\int d^5x\sqrt{-G}\Big[-\frac{1}{4} R+
\frac{1}{2}(\partial\phi)^2+\frac{1}{2}(\partial\chi)^2-
V(\phi,\chi)\Bigr].
\end{equation}
Here, the brane is composed by the two real scalar fields $\phi$ and $\chi$, which depend only on the extra dimension $y$, $R$ denotes the scalar curvature and the space-time is an AdS with $D=5$. Moreover, we adopt the following ansatz for the metric
\begin{equation}\label{act}
ds^2=e^{2A(y)}\eta_{\mu\nu}dx^{\mu}dx^{\nu}+dy^2,
\end{equation}
which preserves the $D=4$ Lorentz invariance \cite{mk1}. The interaction of the two scalar fields modeling the brane will give rise to the warp factor function, which is indicated by $A(y)$. Moreover, we denote the Minkowski space-time metric by $n_{\mu\nu}$ where the indexes $\mu$ and $\nu$ vary from $1$ to $4$.

The equations of motion for the fields can be obtained from the action and might be cast in the form of the following coupled differential equations:
\begin{eqnarray}
\phi'^{2}+\chi'^{2}-2 V(\phi,\chi)&=&6A'^{2}\\\nonumber
\phi'^{2}+\chi'^{2}+2V(\phi,\chi)&=&-6A'^{2}-3A''\\\nonumber
\xi''+4A^\prime\xi'&=& \partial _{\xi}V. \,\,\,\,\,
\end{eqnarray}
In the last equation, $ \xi$ stands for either $\phi$ or $\chi$, while the prime denotes derivative with respect to the extra dimension $y$.

At this point, it is useful to switch to the superpotential method, with which we can obtain first-order differential equations from the motion equations above. This is done by rewriting the potential in terms of a superpotential $W(\phi,\chi)$ \cite{bazeia1, bazeia2, bazeia3}. Accordingly, we obtain for the original potential the following expression
\begin{equation}
V(\phi,\chi)=\frac18 \left[\left(\frac{\partial
W}{\partial\phi}\right)^2 + \left(\frac{\partial
W}{\partial\chi}\right)^2 \right]-\frac13 W^2,
\end{equation}
together with the equations $\phi^{\prime}=\frac12\,\frac{\partial W}{\partial\phi},  \chi^{\prime}=\frac12\,\frac{\partial W}{\partial\chi}$ and $A^\prime=-\frac13\,W$.

In order to obtain the Bloch brane model, one chooses the superpotential function to have the following form \cite{bazeia4, shif,alonso}
\begin{equation}\label{w}
W(\phi,\chi)=2\phi-\frac23\phi^3-2r\phi\chi^2,
\end{equation}
where $r$ is a real parameter that controls the brane thickness.

Finally, we find the following solutions that describe our brane model
\begin{equation}\label{phi}
\phi(y)=\tanh(2ry),
\end{equation}
 \begin{equation}\label{chi}
\chi(y)=\sqrt{\left(\frac1{r}-2\right)}\;{\rm sech}(2ry)
\end{equation} and
\begin{equation}\label{warp}
A(y)=\frac{1}{9r}\Bigl[(1-3r)\tanh^2(2ry)-2\ln\cosh(2ry)\Bigr].
\end{equation}
From the second equation above, we can see that for the limit $r=1/2$, the one-field scenario is
recovered. For certain values of the coupling parameter $r$, there is an splitting of the defect \cite{dionisio}. This brane model supports internal structure which influences the matter energy density along the extra dimension. More details about this feature and graphical analysis of the solutions (\ref{phi}), (\ref{chi}), and (\ref{warp}) are available in the reference \cite{dionisio}.

\section{Gauge field zero mode \label{gau_f_ze_mode}}

Let us now analyze the zero mode of the vector gauge field on the two-field brane model. The basic Bloch brane setup, as well as the other thick brane scenarios, is not capable of supporting the localization of the gauge field zero mode. Indeed, as the warp factor is factorized out of the effective action, one obtains non normalizable solutions from the equations of motion to the gauge field. We fix this problem in the following sections. In order to illustrate these aspects, we start from the action for the gauge field coupled to the gravity as follows
\begin{equation}
S\sim\int d^{5}x\sqrt{-G}F_{MN}F^{MN},
\end{equation}
where the indexes $M$ and $N$ run from $1$ to $5$. The field strength tensor reads  $F_{MN}=\partial_M A_N -\partial_N A_M$.

By choosing the gauge $\partial_\mu A^\mu=A_y=0$, the equation of motion for $A_{\mu}$ is given by
\begin{equation}\label{zero}
-\frac{d^{2}U(y)}{dy^{2}}-2 A'(y)\frac{dU(y)}{dy}=m^{2}e^{-2A(y)}U(y),
\end{equation}
where we have used the ansatz
\begin{equation}\label{sep}
A_\mu(x,y)=a_\mu (0) e^{i p\centerdot x}U(y), \,\,\,\,\,  p^2=-m^2.
\end{equation}
We denote by $x$ the space-time coordinates of the observable universe. The function $U(y)$ possesses all information about the gauge field in the extra dimension, so that we must verify the solutions of the equation (\ref{zero}) in the effective action
\begin{equation}\label{acao}
S\thicksim\int d^{5}x \sqrt{-G}~ F_{MN}F^{MN}= \int dy U(y)^2 \int d^4 xf_{\mu\nu}(x)f^{\mu\nu}(x).
\end{equation}

\begin{figure}
\centering
\includegraphics[scale=1.2]{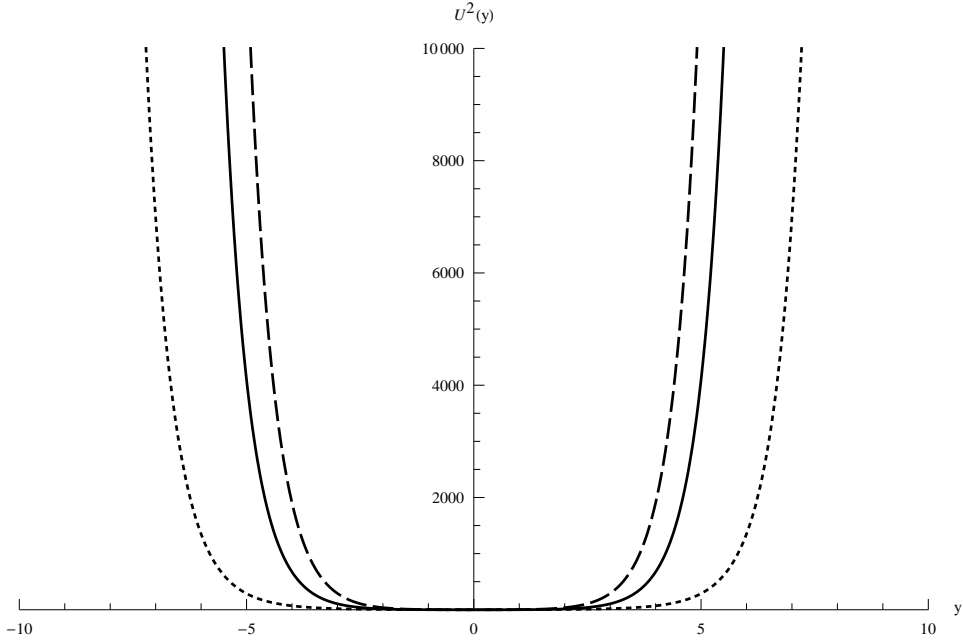}
\caption{{\bf Plots of the solution $U(y)^2$ in equation (\ref{U0}) with $r=0.2$ (points), $r=0.5$ (line) and $r=1$ (dashed line) .}}\label{u0}
\end{figure}
As one sees from the equation above, the warp factor is now absent of the effective action. Therefore, its exponential suppression is lost and the existence of  normalizable zero mode solutions depends on the solutions of the equation of motion to the gauge field in the extra dimension. With $m=0$, we have the solution $U(y)=constant$ and the effective action goes to infinity, so that, with this solution, we can not guarantee the existence gauge field zero mode in this scenario.

Equation (10) has also the solution
\begin{equation}\label{U0}
U(y)=\int_{y_0}^y e^{-2A}dy'.
\end{equation}
However, due to the shape of the function $A(y)$, the solution $U(y)^2$ diverges for any value of $r$. In figure (\ref{u0}) we plot $U(y)^2$ as a function of $y$ for different values of $r$. We note that, $U(y)^2$ diverges for small values of $r$ and, when we increase the value of $r$, the solution diverges more rapidly, which renders the effective action infinite. Therefore, the existence of a localized zero mode with this solution is not guaranteed.

To analyze the massive modes we can write the equation of motion (10) for $m\neq0$ as a Schroedinger-like equation.
\begin{equation}{\bf
\left\{-\frac{d^2}{dz^2}+{V}(z)\right\}\overline{U}(z)=m^2\overline{U}(z)}
\end{equation}
with the potential given by
\begin{equation}\label{v0}{\bf
V(z)=-\frac{7}{4}\dot{A}^2+\frac{1}{2}\ddot{A}}
\end{equation}
This equation cannot be written in the form corresponding to supersymmetric quantum mechanics $Q^{\dag}\,Q\,\overline{U}(z)=-m^2\overline{U}(z)$. For this reason, we cannot exclude the possibility of tachyonic states in the spectrum \cite{dionisio_tachyons}.

\begin{figure}
\centering
\includegraphics[scale=1]{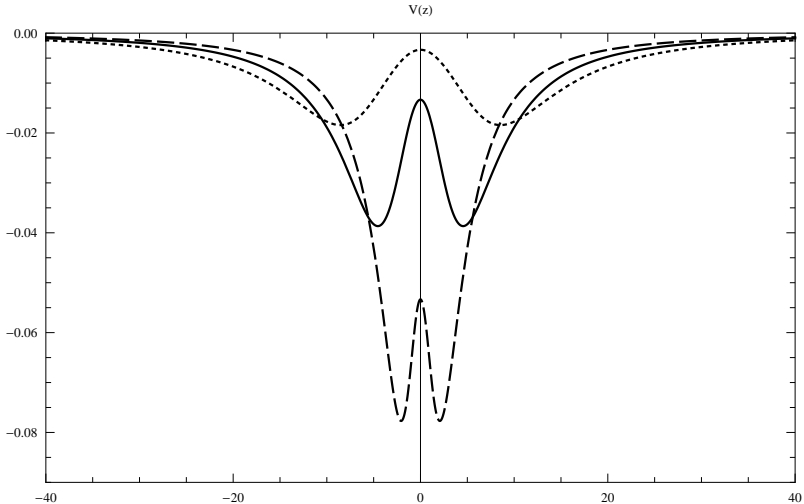}
\caption{ Plots of the potential $V(z)$ in equation (\ref{v0}) with $r=0.05$ (points), $r=0.1$ (line) and $r=0.2$ (dashed line) .}
\label{potzero}
\end{figure}

The absence of resonant modes can be verified by the structure of the resulting potential, which we show in figure (\ref{potzero}). As we have noted the potential tends to infinity with negative values, and the maximum values are also negative. In addition, we note that the potential does not have the usual volcano-like structure which has been obtained in thick brane models with localized zero modes as well as gauge field resonant modes.

As we have remarked, the standard Bloch brane background is not capable to supporting the existence of a localized zero mode of the gauge field on the brane. Following recent results, we modify the two-field brane model presented in Section \ref{gau_f_ze_mode} in search for better prospects for localizing gauge fields in thick branes.

\section{Dilatonic Bloch Brane \label{dil_bb}}

In order to find a way around the difficulties for localizing gauge field on the brane model presented in Section \ref{gau_f_ze_mode}, we apply a method largely used to implement the localization of fields in thick brane setups \cite{kehagias, mk1, nosso1,mk2, mk3, nosso2}. In this way, we study the features of the vector gauge field in a scenario we find convenient to call the dilatonic Bloch brane. It is described by the action
\begin{equation}\label{acdil}
\mathcal{S}=\int d^5x\sqrt{-G}\left(-\frac{1}{4}R+\frac{1}{2}(\partial\phi)^2+\frac{1}{2}(\partial\chi)^2+\frac{1}{2}(\partial\pi)^2-V(\phi,\chi,\pi)\right).
\end{equation}
Following the reference \cite{kehagias}, we keep the scalar fields $\phi$ and $\chi$, which describe the brane. Then, we include the dilaton scalar field $\pi(y)$ that will couple directly with the gauge field. This is the key point to achieve normalizable solutions.

In the following we adopt the ansatz for the metric \cite{kehagias}
\begin{equation}
ds^2=e^{2A(y)}\eta_{\mu\nu}dx^{\mu}dx^{\nu}+e^{2B(y)}dy^2.
\end{equation}
Then, the resulting equations of motion from the corresponding action read
\begin{eqnarray}
\phi'^{2}+\chi'^{2}+\pi'^{2}-2e^{2B}V&=&6A'^{2}\\\nonumber
\phi'^{2}+\chi'^{2}+\pi'^{2}+2e^{2B}V&=&-6A'^{2}+3A'B'-3A''\\\nonumber
\gamma''+(4A'-B')\gamma'&=& e^{2B}\partial _{\gamma}V, \,\,\,\,\, \gamma=\phi,\chi,\pi.
\end{eqnarray}
At a first glance, it is not clear which modifications are brought about by the dilaton field, but this will be clarified when we complete the background description. In order to obtain first-order differential equations we use the same superpotential method applied in Section \ref{rev_bb_sce}.  With the aim of keeping the Bloch brane structure we consider the following potential function \cite{kehagias}
\begin{equation}
V=e^{\pi \sqrt{\frac{2}{3}}}\left[\frac{1}{8}\left(\frac{\partial W}{\partial \phi}\right)^2+\frac{1}{8}\left(\frac{\partial W}{\partial \chi}\right)^2 -\frac{5}{16}W^2\right].
\end{equation}
We obtain first-order differential equations that solve the equations of motion
\begin{eqnarray}\label{singularidade1}
\phi^{\prime}=\frac12\,\frac{\partial W}{\partial\phi},\,\,\,\,\,
\chi^{\prime}=\frac12\,\frac{\partial W}{\partial\chi}\\\nonumber
\pi=-\sqrt{\frac{3}{8}}A,\\\nonumber
B=-\frac{\pi}{2}\sqrt{\frac{2}{3}}=\frac{A}{4},\\\nonumber
A'=-\frac{W}{3}.
\end{eqnarray}
Choosing the same superpotential function (\ref{w}), the solutions to $\phi$, $\chi$ and $A(y)$ displayed in equations (\ref{phi}), (\ref{chi}), and (\ref{warp}) remain valid in the present case.

In view of this new background, let us turn our attention to the effective action for the gauge field. We introduce the coupling with the dilaton as follows \cite{kehagias, mk1, nosso1,mk2, mk3, nosso2}
\begin{equation}
S\sim\int d^{5}x\sqrt{-G}~e^{- 2\lambda \pi\sqrt{\frac{2}{3}}}F_{MN}F^{MN},
\end{equation}
where $\lambda$ represents a parameter which controls the dilaton coupling.

Again, using the gauge $\partial_\mu A^\mu=A_y=0$ together with the ansatz $A_\mu(x,y)=a^\mu (0) e^{i p\centerdot x}U(y)$ and $p^2=-m^2$, we obtain the equation for part of the gauge field which depends on the extra dimension, namely
\begin{eqnarray}\label{zero2dil}
-\frac{d^{2}U(y)}{dy^{2}}-\left[2 A'-B'-2\lambda \pi'\sqrt{\frac{2}{3}}\right]\frac{dU(y)}{dy}=m^{2}e^{2(B-A)}U(y).
\end{eqnarray}
From the equation above, for $m=0$ and using the relations (\ref{singularidade1}) we have the solution $U(y)=constant$, that is enough to guarantee the existence of gauge field zero mode localized on the brane. In this way we take the effective action and decompose it in the following way
\begin{equation}{\bf
\int d^{5}x \sqrt{-G}~e^{- \lambda \pi\sqrt{\frac{8}{3}}}F_{MN}F^{MN}= \int dy U(y)^2 e^{A(y)(\frac{1}{4}+\lambda)} \int d^4 xf_{\mu\nu}(x)f^{\mu\nu}(x),}
\end{equation}
where we have used again the proportionality relation between the functions $\pi(y)$ and $A(y)$, as in eqs. (20). We remark that, in order to recover the background without the dilaton coupling we must, in addition to suppressing the coupling $\lambda$, return to the metric ansatz (2), proceeding in an analogous manner as in references \cite{mk1,kehagias, nosso1, mk2, nosso2}.

Direct comparison of the action above with the action (\ref{acao}), which does not include the dilaton, shows that the inclusion of the dilaton on the brane model brought back the warp factor for effective action. We can see clearly that for $\lambda>-1/4$ the solution is normalizable. The dilaton coupling has the effect of limiting the range of the gauge field in the extra dimension, a desired feature when one wishes to achieve localization. Thus, we conclude that the dilatonic Bloch brane does support the existence of gauge field zero modes.

\section{Massive modes and resonances \label{mass_mode_res}}

In this section, we address the question of how the massive modes interact with the dilatonic brane scenario. In order to analyze the Kaluza-Klein massive spectrum, we return to the equation (\ref{zero2dil}) with $m\neq0$ and transform it into a Schroedinger-like equation. This can be carried out by using the prescription
\begin{equation}\label{t}
y\rightarrow z=f(y),\,\,\,\,\,\,f'(y)=e^{-\frac{3}{4}A}\,\,\,\,\,\, U(y)=e^{-\gamma A}\overline{U}(z),
\end{equation}
where $\gamma=(\lambda+1)/2$. In this way we arrive at the following
Schroedinger-like equation
\begin{equation}\label{schgauge}
\left\{-\frac{d^2}{dz^2}+{V}(z)\right\}\overline{U}(z)=m^2\overline{U}(z),
\end{equation}
with the one-particle potential $V(z)=\gamma (\gamma\dot{A}^2+\ddot{A})$. Here, the functions $\dot{A}$ and $\ddot{A}$ represent, respectively the first derivative and the second derivative with respect to $z$. Equation (\ref{schgauge}) can also be written as the form corresponding to supersymmetric quantum mechanics
\begin{equation}\label{susy_qm_gauge}
Q^{\dag} \, Q \,
\overline{U}(z)=\left\{\frac{d}{dz}+\gamma\dot{A}\right\}\left\{\frac{d}{dz}-\gamma\dot{A}\right\}\overline{U}(z)=-m^2\overline{U}(z).
\end{equation}
This excludes the possibility of tachyonic states in the spectrum \cite{dionisio_tachyons}.

\begin{figure}
\centering
\includegraphics[scale=1.55]{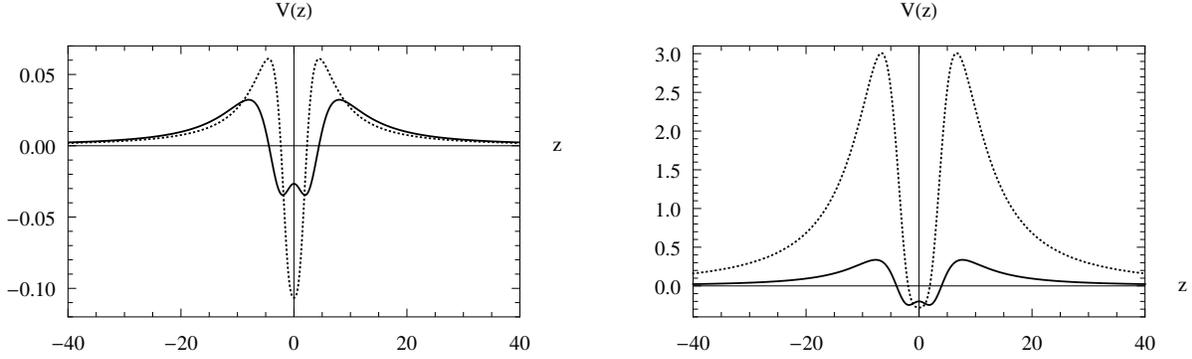}
\caption{LEFT: Plots of the potential
$V(z)$ for $[r=0.1,\lambda=1]$ (solid line), $[r=0.2,\lambda=1]$ (points). RIGHT: Plots of $5V(z)$ for $[\lambda=1,r=0.1]$ (solid line) and $V(z)$ for $[\lambda=20,r=0.1]$ (doted line). Notice that, in order to optically accommodate both graphs in a single frame, the potential in the graph for $[\lambda=1,r=0.1]$ appears multiplied by a factor of $5$.}
\label{pot1}
\end{figure}

In order to investigate the existence of gaps in the spectrum, we must study the asymptotic behavior of the potential $V(z)$. However, due to the transformations (\ref{t}) we can not find the function $A(z)$ analytically, which means that the potential must be studied numerically. The potential obtained in this way is shown in Figure (\ref{pot1}). In the left-hand side of Figure (\ref{pot1}), we observe the influence of parameter $r$ on the potential. As we can see, when $r$ tends to zero the minimum at $z=0$ splits into two minima, revealing the existence of internal structure as noticed in reference \cite{dionisio}. In the present article, we are more interested in the influence of the dilaton coupling on massive modes. For this reason, on the right side of Figure (\ref{pot1}), we focus on the $\lambda$-dependence of the potential. Notice that increasing the intensity of the dilaton coupling (raising the value of $\lambda$), the values of the two maxima of the potential increase while the two minima are merge into a single one. This feature is very interesting, considering that it changes the behavior of the massive modes in the region between the maxima, i.e., in the region where the four-dimensional universe is located. Therefore, the dilaton coupling will contribute to the interaction of the massive modes with the brane.

\begin{figure}
\centering
\includegraphics[scale=1.55]{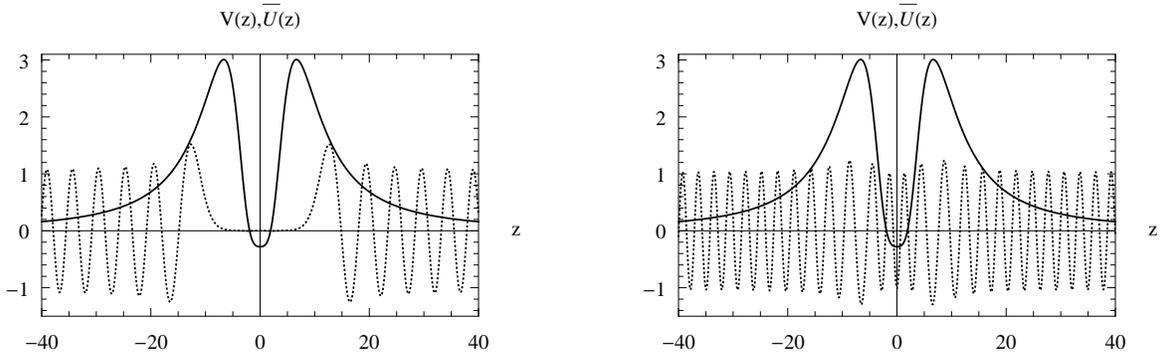}
\caption{Plots of $V(z)$ (solid line) and $\overline{U}(z)$ (points) for $r=0.1$, $\lambda=20$, $m=1$ (left), $m=3$ (right).}\label{pot2}
\end{figure}

Let us now turn to the the effects on the the solutions of the Kaluza-Klein (KK) modes that arise due to the increasing of the maximum values of the potential relative to the square mass. We present in Figure (\ref{pot2}) the plots of the potential $V(z)$ and of the function $\overline{U}(z)$ for the mass values $m=1$ and $m=3$.
As we can see, light modes are suppressed in the region between the two maxima, however for $m^2> V_{max}$ the wave function oscillates in the region near the brane. Observing the asymptotical behavior of the potential we conclude that
\begin{equation}\lim _{z \rightarrow \infty}{\{V(z)\}} =0,
\end{equation}
and the continuum spectra is free of gaps.

In the case of very large masses, satisfying the condition $m^2>>V_ {max}$, the potential represents only a slight perturbation. Therefore, resonant modes would rather be expected in the region $m^ 2<V_ {max}$ \cite{gremm}. Indeed, as noticed in references \cite{gremm,csaba,csaba2}, for some specific masses, the plane wave solutions of the Schroedinger equations can resonate with the potential, thus exhibiting exceedingly large amplitude inside the brane. In the present case, such structures will show us how the gauge field massive modes couple to matter on the brane.

In order to search for the existence of resonant modes we will consider the probability density of the wave function at the center of the brane, an idea which has been applied successfully before \cite{num,cvetic,chineses1,chineses2,chineses3,chineses4,nosso1,nosso2,nosso3,nosso4}. To this end, we evaluate the probability density of the normalized wave function at $z=0$ for a given value of $m$, which is defined according to
\begin{equation}
P(m)=\frac{|\overline{U}_m(0)|^2}{\int_{-200}^{200}|\overline{U}_m(z)|^2 dz}.
\end{equation}

\begin{figure}
\centering
\includegraphics[scale=0.62]{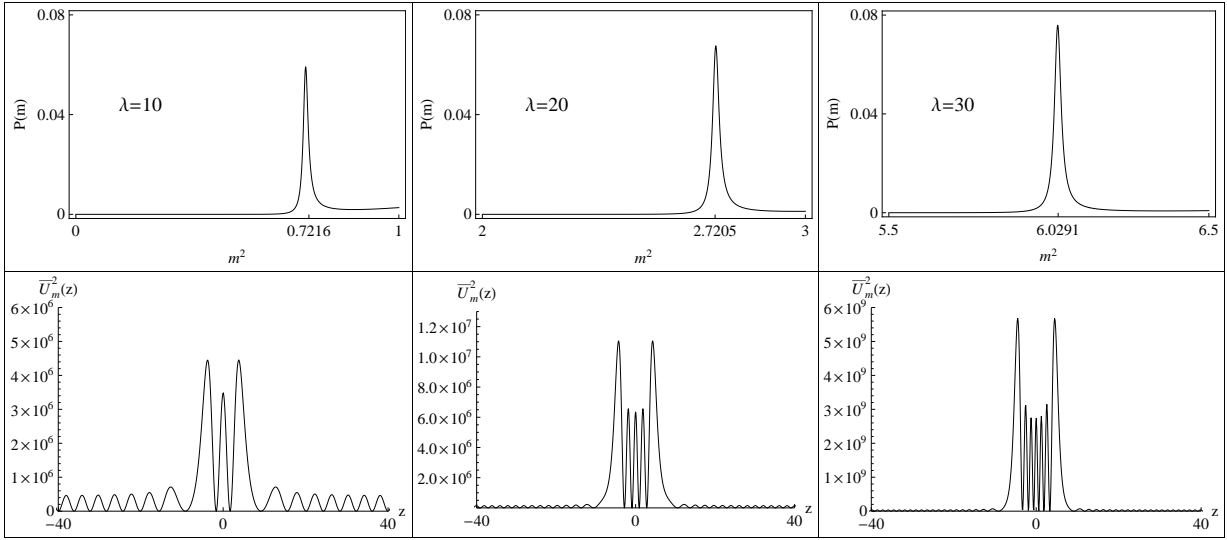}
\caption{Plots of $P(m)$ (up) and $\overline{U}^2(z)$ (down) for $m^2=0.7216$ (left), $m^2=2.7205$ (center) and $m^2=6.0291$ (right). We have used $r=0.1$.}\label{reso}
\end{figure}

In the top of Figure (\ref{reso}) we show $P(m)$ for some values of the coupling constant $\lambda$. For $\lambda=10, 20, 30$ we find resonant modes respectively for masses $m^2=0.7216, 2.7205, 6.0291$. To confirm that our technique is suitable for detecting resonant modes, in the Figure (\ref{reso}), we showed the solutions of the wave functions for the values of masses indicated with peaks in $P(m)$. We note that these specific energies correspond to wave functions with very high amplitudes inside the brane.

It is remarkable that we find heavy resonant modes. This is in contrast to what we have observed previously in the case of Kalb-Ramond fields \cite{nosso1} and of a gauge field \cite{nosso2}, both in deformed branes, and also of fermions in thick branes \cite{nosso3} and also of graviton resonances \cite{nosso4}, where the resonant modes are only found for very low values of the mass. We attribute this effect to the increasing of the coupling between the gauge field and the dilaton. A similar feature has been reported in \cite{mk2}.

\begin{figure}
\centering
\includegraphics[scale=0.7]{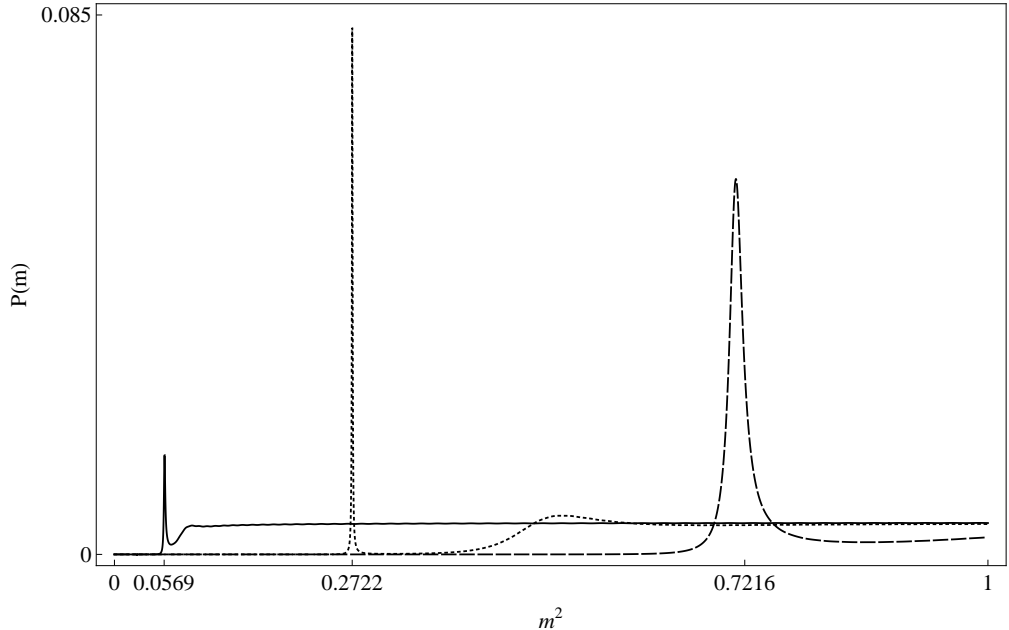}
\caption{Plots of $P(m)$ for $\lambda=10$, $r=0,01$ (solid line), $r=0.05$ (points)  and $r=0.1$ (dashed line).}\label{reso2}
\end{figure}

Another important point to be considered is about the influence of the brane internal structure on the resonant modes. As we know, the thickness of the defect increases with decreasing $r$ \cite{dionisio,adauto}. This effect results in an increased spacing between the two maxima of the gauge field potential, as noted in Figure (\ref{pot1}). To analyze how this change in the internal structure interferes in the coupling of massive modes with the brane, we show in Figure (\ref{reso2}) the function $P(m)$ keeping  $\lambda$ fixed and reducing the value of $r$. We conclude that reducing the thickness of the defect causes an effect similar to increasing the dilaton coupling, i.e., the resonances occur for heavy masses.

\begin{figure}
\centering
\includegraphics[scale=1.5]{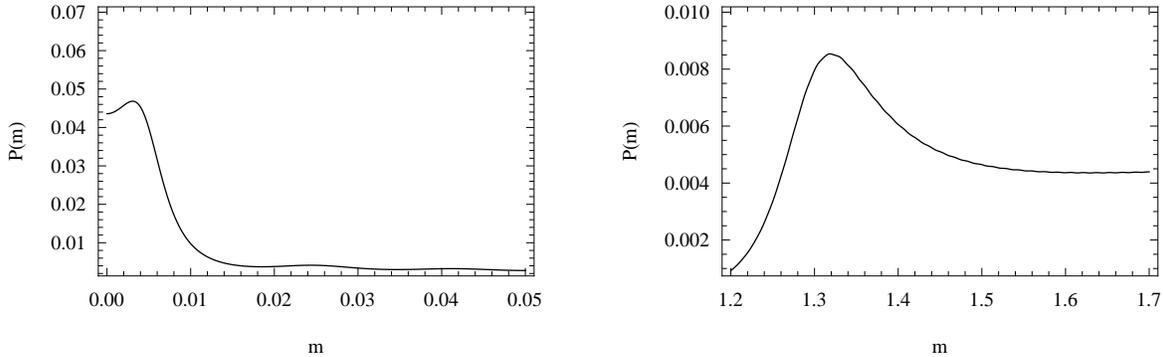}
\caption{{\bf Plots of $P(m)$ with $[r=0.1;\lambda=0]$ on left and  $[r=0.2;\lambda=10]$ right. We note the presence of broader peaks.}}\label{lifetime}
\end{figure}

Observing figures (\ref{reso}) and (\ref{reso2}) we verify that the masses of the resonant modes depend on the values of $r$ and $\lambda$.  Increasing the value of these constants leads to resonant modes with larger masses. However for certain combinations of these two constants we obtain very broad peaks for the $P(m)$. This characterizes states with life-time too short to result in any physical effect. The life-time $\tau$ of observed resonances can be estimated by $(\Delta m)^{-1}$ where $\Delta m$ is the width in mass at half maximum of the peaks in $P(m)$ \cite{carlos, chineses1, lifetime, chineses2, chineses3}. In figure (\ref{lifetime}) we plot two examples of broad peaks in $P(m)$ with $\Delta m > m$ that cannot be considered as resonant states. As a rule, for small values of $\lambda$ or for values of $r$ approaching to $0.5$, we obtain structures with too short life-times ($\Delta m > m$). On the other hand, for larger values of $\lambda$ and for lower values of $r$, we obtain longer life-times. Furthemore, when we find peaks on the $P(m)$ function that may be considered as resonances, they appear only once in the mass spectrum for each potential considered in the equation of motion. This interesting aspect of finding more than one resonance for the same potential, for instance, has been found in another article, where we have considered the study of fermions in deformed branes \cite{nosso3}.

\section{Conclusions \label{conclu}}

We have analyzed several aspects of the localization properties of the vector gauge field in a brane generated by two scalar fields coupled to gravity: the Bloch brane. Initially, by considering the basic defect setup we have concluded that there is no normalizable solution for the zero mode of the gauge field. As the warp factor does not appear in the effective action, there is no convergence factor and the solutions of the equations of motion render the part of the effective action regarding the extra dimension infinite. To circumvent this problem, we have applied a method widely used in the study of vector and tensor gauge fields in thick membrane models. By introducing the dilaton field to the background, and thereby bringing the exponential suppression from the warp factor back to the effective action, we were able to guarantee the existence of a localized zero mode.

This new scenario allowed us to analyze the massive spectrum in a supersymmetric quantum mechanics framework. In this way, we have transformed the equations of motion into a Schroedinger-like equation where we have analyzed the influence of the potential on the wave solutions. The potential extracted from the Schroedinger equation exhibits a splitting caused by the internal structure of the membrane. When we increase the intensity of the dilaton coupling, the values of the two maxima of the potential are increased while the two minima merge into one another. Thus, when we increase $\lambda$ the coupling with the membrane is suppressed for all but the heavy modes. This feature stems from the fact that, for masses obeying $m^2 >> V_{max}$, the potential represents only a small perturbation.

We have found new resonance structures in the massive spectrum. These new resonant modes are different and posses the interesting property that, contrary to what is found in previous similar investigations, they appear also for large masses depending on the intensity of the dilaton coupling. By considering the probability density of finding the modes at $z=0$, we have found a variety of such resonant modes, which carry information about the coupling of massive modes with matter on the membrane in terms of the coupling with the dilaton field and the internal dynamics of the membrane. Regarding the internal structure of the membrane, we have shown that reducing the thickness of the defect increases the value of the masses in which the resonances occur. Similarly, when we increase the value of the constant-coupling $\lambda$, we also obtain resonant modes for larger masses. This shows that the coupling of the dilaton is able to favor the coupling with the membrane with heavy modes.

The authors would like to thank the Funda\c{c}\~{a}o Cearense de apoio ao Desenvolvimento
Cient\'{\i}fico e Tecnol\'{o}gico (FUNCAP), the Coordena\c{c}\~ao de Aperfei\c{c}oamento de Pessoal de N\'ivel Superior (CAPES), and the Conselho Nacional de Desenvolvimento Cient\'{\i}fico e Tecnol\'{o}gico (CNPq) for
financial support.  The author W. T. Cruz would like to thank to the C0R3 CLAN friends for the excellent discussions. One of us (A. R. P. L.) would like to acknowledge the hospitality of the department of Physics of the Federal University of Cear\'a, where he spent part of 2012 as a postdoc.

\end{document}